\documentclass[twocolumn,showpacs,amsmath,amssymb,pre,aps]{revtex4}   
\usepackage{graphicx}
\usepackage{dcolumn}
\usepackage{bm}
\usepackage{color}
\usepackage{hyperref} 
\renewcommand{\vec}[1]{\mathbf{#1}}

\newif\ifgraph

\graphtrue             %

\begin{document}
\title{Directed autonomous motion of active Janus particles induced by wall-particle alignment interactions}

\author{Poulami Bag$^1$, Tanwi Debnath$^2$, Shubhadip Nayak$^1$ and Pulak K. Ghosh$^1$}
\email{pulak.chem@presiuniv.ac.in}

\affiliation{ Department of Chemistry,
	Presidency University, Kolkata 700073, India}

\affiliation{ Theoretical Physics of Living Matter, Institute for Advanced Simulation, Forschungszentrum J\"ulich, 52425 J\"ulich, Germany}

\date{\today}

\date{\today}

\begin{abstract}
We propose a highly efficient mechanism to rectify the motion of active particles by exploiting particle-wall alignment interactions. Through numerical simulations of active particles' dynamics in a narrow channel, we demonstrate that a slight difference in alignment strength between the top and bottom walls or a small gravitational drag suffices to break upside-down symmetry, leading to rectifying the motion of chiral active particles with over 60\% efficiency. In contrast, for achiral swimmers to achieve rectified motion using this protocol, an unbiased fluid flow is necessary that can induce orbiting motion in the particle's dynamics. Thus, an achiral particle subject to Couette flow exhibits spontaneous directed motion due to an upside-down asymmetry in particle-wall alignment interaction. The rectification effects caused by alignment we report are robust against variations in self-propulsion properties, particle's chirality, and the most stable orientation of self-propulsion velocities relative to the walls. Our findings offer insights into controlled active matter transport and could be useful to sort artificial as well as natural microswimmers (such as bacteria and sperm cells) based on their chirality and self-propulsion velocities.
  
\end{abstract}
  \maketitle

\section{Introduction} \label{intro}

Active particles are motile entities that can move independently by extracting energy from their environment~\cite{Granick,Muller,Paxton,Sen,Wang,Romanczuk1,Marchetti,Bechinger,Gompper}. They exist in various sizes and can be either natural or artificial. Depending on their design and intended functions, artificial active particles come in many shapes and are referred to by different names. One of the most significant categories of active particles is the self-propelled Janus particle~\cite{Granick,Muller,Paxton,Sen,Wang,misko-1,misko-2,misko-3}, characterized by having two distinct faces with different physical and chemical properties, which grants them unique functionalities for autonomous motion.

In recent years, artificial active particles have garnered considerable attention in research due to their potential applications in nanotechnology and biotechnology~\cite{ap1,ap2,ap3,ap4,ap5,ap6,ap7,ap8,ap9}. Prominent uses include targeted drug delivery~\cite{ap6}, cargo transport~\cite{ap7}, chemotherapy~\cite{cancer-therapy1}, and cell therapy~\cite{cell-therapy1,cell-therapy2}. Beyond therapeutic uses, active particles could assist in monitoring the movements of slower particles, illuminating previously elusive biological processes~\cite{ap4,nanoscale,photo-system-1,photo-system-2}. Further, they can act as sensitive detectors for biological and chemical agents, providing valuable insights across various fields~\cite{ap8,ap9}. Their potential applications in imaging technologies~\cite{imaging1,imaging2}, particularly in optical nanoscopy~\cite{nanoscopy} and intraocular surgery~\cite{intraocular-surgery-1,intraocular-surgery-2}, along with their versatility and promise, position active particles as a cornerstone for future scientific advancements.

Beyond their wide-ranging technological utilities, active particles serve as an intriguing model system that exhibits non-equilibrium phenomena of fundamental importance~\cite{Romanczuk1,Marchetti,Bechinger,Gompper,funda-1,funda-2,MSshort,our-review,Daisuke,misko-4,misko-5,physics-fluid-3,physics-fluid-4}. For example, active particles demonstrate autonomous directional motion when subjected to a spatial periodic structure that lacks inversion symmetry, a phenomenon known as ratcheting~\cite{MSshort,our-review,ratchet,Ai-rat,Reichhardt1,JPCL,physics-fluid-2}. They also display apparent drift against an external bias~\cite{GNM,Ai-negative,Reichhadt2,physics-fluid-1}, transiently move toward stimuli~\cite{stimuli-1,stimuli-2,stimuli-3}, and notably, undergo motility-induced phase separation~\cite{ Fily,Redner1,Redner2,Omar_Brady,Nardini,Tailleur,Tailleur2,Loewen1,3D1,3D2,Lowben-e1,Nayak_mips,
	Castro1,Castro2}.

However, the unique dynamical behavior of active particles presents challenges for controlling their transport, which limits their applications in various fields. A critical issue is how to direct the motion of these particles. Despite their persistent movement, over a long time scale (typically much longer than the rotational relaxation time), particles tend to exhibit random motion. A well-known approach to solving this problem is to rectify their motion using the ratchet effect~\cite{hanggi-artf,Erbas}. According to Pierre Curie’s conjecture, an active particle subjected to spatial structures with broken inversion symmetry can exhibit autonomous directed motion. Various ratcheting protocols have been proposed to achieve the rectification of active particle motion, often involving periodic channels, substrate potentials, or fluid flow patterns that create a periodic structure. In this work, we demonstrate a new mechanism that does not rely on any periodic structure.

When an active particle diffuses through a narrow channel, the particle's self-propulsion velocity direction is likely to be affected due to particle-wall alignment interaction~\cite{Uspal,Bianchi,Mozaffari,Podda,Czajka,Jalilvand}. In addition to the particle-wall hydrodynamic interactions, the alignment effect in active Janus particles may arise from factors like~\cite{PBag2024}: (i) the differing dielectric properties of the materials in the particle's hemispheres, which can lead to stabilizing interactions at specific orientations, and (ii) the unequal exposure of the coated surface to the fuel near the confining boundary, resulting in varying self-phoretic rates across the hemispheres. This difference can influence the self-propulsion direction as the particles approach the wall. Previous studies show that particle-wall interaction greatly influences the dynamics of particles~\cite{binous,mangeat, Boymelgreen,m-li}, inducing rich phenomenology. Examples include stationary particle currents in sedimenting active matter wetting a wall\cite{mangeat}, separation of Janus droplets and oil droplets~\cite{m-li}, and particle-electrode wall interactions in manipulating the mobility of active Janus particles\cite{Boymelgreen}. 

In this work, we demonstrate how the interaction between particles and walls can be utilized to guide randomly moving active particles in a desired direction while controlling their average velocity. Our study reveals that a slight difference in the alignment-induced coupling at the top and boundary walls, or a small gravitational drag due to the apparent weight of the particles, is sufficient to break the symmetry of a straight channel. This spatial structure, characterized by broken inversion (upside-down) symmetry, enables the autonomous motion of chiral active particles to be directed toward a specific direction with a significantly high rectification power. We observe a large rectification effect over a wide range of self-propulsion and alignment interaction parameters. The amplitude and direction of the average particle current largely depend on the direction and magnitude of the chiral torque. Moreover, we explore a possible way for rectifying the motion of achiral particles by exposing them to an unbiased Couette flow combined with particle-wall alignment interaction.

Outlays of this paper are as follows: In Sec.~II, we describe the model of particle dynamics and the particle-wall alignment interaction. We also discuss the meaning and significance of all controlled parameters. In Sec.~III, we present our key numerical results and provide some analytical arguments, conducting a detailed analysis of the autonomous directed motion of both chiral and achiral particles under various stable configurations. Finally, in Sec.~IV, we summarize our findings and offer concluding remarks.


\section{Model}\label{Model}
\begin{figure}[t]
	\centering	\includegraphics[width=0.45\textwidth,height=0.2\textwidth]{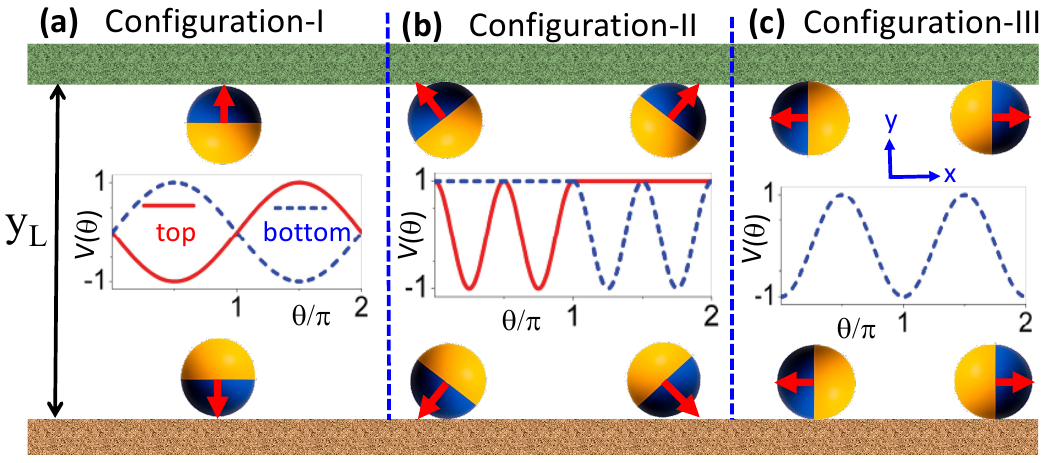}
	\caption {(color online) A schematic illustration depicts various stable self-propulsion velocity orientations, shown by red arrows. The associated interaction potential $ V(\theta) $ [refer to Eqs.~(\ref{top_V})] is represented by solid and dotted lines for the top and bottom walls, respectively. (a) Configuration I: The particle is most stable at angles \( \theta = {\pi}/{2} \) (near the top wall) and \( \theta = {3\pi}/{2} \) (near the bottom wall). For this configuration, the interaction potential with the top wall is given by Eq. (\ref{top_V}), where \( {q} = 1 \) and \( \phi = 0 \). For the bottom wall the alignment interaction potential,  $V_b(\theta) = -\kappa V_t(\theta)$.
		(b) Configuration II: In this case, the stable orientation of \( \vec{v_0} \) occurs at \( \theta = {\pi}/{4} \) and \( \theta = {3\pi}/{4} \) near the top wall. Near the bottom wall, the stable orientations correspond to \( \theta = {5\pi}/{4} \) and \( \theta = {7\pi}/{4} \). For this configuration, \( {q} = 4 \) and \( \phi = 0 \). Note that for the top wall, \( V(\theta) \) is given by Eq. (\ref{top_V}) for \( 0 \leq \theta \leq \pi \); otherwise, \( V(\theta) = 0 \). For the bottom wall,  $V_b(\theta) = \kappa\omega(y) \cos(4\theta)$ for \( \pi \leq \theta \leq 2\pi \); otherwise, \( V(\theta) = 0 \).
		(c) Configuration III: Here, the particle-wall interaction energy reaches its minimum when \( \vec{v_0} \) is parallel to the wall.  The interaction potential is described by Eq.~(\ref{top_V}), with \( {q} = 2 \) and \( \phi = {\pi}/{2} \). Further, for this configuration, the interaction potential for both the top and bottom wall alignments is the same for $\kappa =1$, [that is $V_b(\theta)=\kappa V_t(\theta)$ ].}
	\label{F1}
\end{figure}
We consider an active Brownian particle of circular disk that is diffusing in a two-dimensional (2D) channel filled with a highly viscous fluid. To avoid unessential complications, we restrict our analysis to the case of 2D channel. We anticipate that most of the findings can be easily extended to 3D channels with flat walls. However, for complicated surface structures, the interaction between particle-wall alignment interaction would be complex, and the results in 2D could be significantly different from the 3D cases.

We assume the particle is a hard disk, and any collisions with the walls of the channel are perfectly elastic.   Beyond these collisional interactions, the particle has a tendency to align preferentially against the boundary walls. The dynamics of the particle’s center of mass, denoted by coordinates $(x, y)$, are described by the following set of overdamped Langevin equations:
\begin{eqnarray}
	\dot{x} &=& v_0 \cos{\theta} + u_{s}(y) + \sqrt{2D_0}\;\xi_{x}(t)\label{L1} \\
	\dot{y} &=& v_0 \sin{\theta} - g + \sqrt{2D_0}\;\xi_{y}(t)\label{L2}\\
	\dot{\theta} &=& \Omega_I + \Omega_{s}(y) + \Omega_w + \sqrt{2D_{\theta}}\;\xi_\theta(t)\label{L3}
\end{eqnarray} 
Where $v_0$ is the amplitude of self-propulsion velocity, with components ($v_0 \cos\theta, v_0 \sin\theta$). The angle $\theta$ denotes the direction of the self-propulsion velocity with respect to the channel axis (x-axis).

When an active particle is subjected to a shear flow, it experiences both advection drag,  $\gamma_t u_s(y)$, and shear torque, represented as $\gamma_r \Omega_s(y)$. Here, \(\gamma_t\) and \(\gamma_r\) are coefficients representing the frictional forces (and torque) associated with translational and rotational motion, respectively. As a  study case, we consider an unbiased Couette flow. In Sec.~IIIC, we provide a brief overview of its velocity profile [see   Fig.~\ref{F6}(a)]. Further, the particle experiences a drag $g \gamma_t$ perpendicular to the channel axis due to its apparent weight (i.e., weight minus buoyant force).

The last terms, $\xi_{x}(t)$, and $\xi_{y}(t)$ of Eq.~(\ref{L1}-\ref{L2}) represent thermal fluctuations; they are characterized by a Gaussian distribution and the following statistical properties.

\begin{eqnarray}
	\langle \xi_{x}(t) \rangle &=&\langle \xi_{y}(t) \rangle = 0,  \nonumber \\
	\langle \xi_{x}(t) \xi_{x}(t') \rangle &=& \langle \xi_{y}(t) \xi_{y}(t') \rangle  = \delta({t-t'}).
	\label{noise}  
\end{eqnarray}
Here,  $D_0$ is the thermal noise strength associated with translational motion, as well as the diffusion constant of a free Brownian particle when $v_0=0$ and the fluid is at rest. For a spherical particle of radius $r_0$, $D_0$ is related to the temperature ($T$) and viscosity ( $\eta$) of the medium via the Stokes-Einstein relation, $D_0 = k_B T/6\pi\eta r_0=k_B T/\gamma_t$. Where $k_B$ is the Boltzmann constant. Note that this expression for the diffusion constant is based on the assumption that the momentum flow of the suspension fluid is not confined to a single plane. Therefore, calculating the relevant hydrodynamic friction is inherently a three-dimensional task. As a result, our model can be considered as a quasi-two-dimensional, as previously utilized in many works~\cite{quasi-2D-1,quasi-2D-2,quasi-2D-3}.


The time evolution of the direction of self-propulsion velocity [see Eq.~(\ref{L3})], with respect to laboratory x-axis, is governed by intrinsic angular velocity ($\Omega_I$), fluid flow induced angular velocity ($\Omega_s$), and particle-wall alignment interaction induced angular velocity ($\Omega_w$), in addition to the rotational diffusion.   The last term in Eq.~(\ref{L3}), $\xi_{\theta}(t)$, governs rotational diffusion  $D_\theta$ of the active particle, exhibits similar statistical properties to the thermal translational noise terms $\xi_{x}(t)$ and $\xi_{y}(t)$. In free space, the rotational diffusion of a colloidal particle can be estimated using the Einstein-Smoluchowski relation, $D_{\theta}=k_B T/8\pi\eta r_0^3=k_B T/\gamma_r$. The rotational relaxation time and self-propulsion length are expressed by $\tau_\theta = 1/D_\theta$ and $l_\theta = \tau_\theta v_0$, respectively. For active particles, due to the self-propulsion mechanism, the particle orientation varies in a more complex way. Therefore, in our study, we consider the rotational and translational diffusion constants to be independent parameters.

In the overdamped limit, torque is proportional to angular velocity, and force is proportional to linear velocity. Thus, our discussion often refers to angular velocity as torque and linear velocity as force or drag.

{\it Particle-wall alignment interaction} --- As demonstrated in ref.~\cite{PBag2024}, particle-wall alignment interaction can cause active particles to preferentially assume a specific orientation  with respect to the wall. Such a type of interaction affects active particles' dynamics and hence diffusion, as well as transport properties when they are close to the boundary walls. To model particle-wall alignment interaction, we assume that when the particle gets closer than a certain cut-off distance $\lambda$, the alignment interaction-induced torque starts affecting particle dynamics. The strength of this interaction decays exponentially as the particle moves away from the wall. 
\begin{eqnarray}
	\omega(y) &=& \gamma_r \omega_a \exp[-\chi |d(y)-\sigma/2|], \; \; \; {\rm if} \; \; d(y) \le \lambda. \nonumber \\
	&=& 0 \; \; \; \;   {\rm otherwise} 
	\label{w}
\end{eqnarray}
Here, $1/\chi$ has similar significance as Debye length, and $\sigma$ is the particle's diameter. The alignment interaction strength assumes its maximum value $\gamma_r \omega_a$ at the separating distance, $ d(y)=\sigma/2$.  
The alignment interaction-induced angular velocity ($\Omega_{w}$) is derived from the orientation-dependent potential, $\Omega_{w} = -(1/\gamma_r)\partial V (\theta)/\partial \theta$.  Where, $V (\theta)$ stabilizes specific orientations of the particle's self-propulsion velocity with respect to the wall. For
quantitative analysis, we choose the following form the $\theta$ dependent alignment potential with the top ($V_t$) wall of the channel,
\begin{eqnarray}
	V_t(\theta) = -\omega(y)\sin({q}\theta + \phi)
	\label{top_V}
\end{eqnarray}
where $\phi$ is the phase factor, ${q}$ represents the number of stable degenerate configurations, and $\omega(y)$ is the distance-dependent strength of the alignment interaction [see Eq.~(\ref{w})]. The parameters, $\phi$ and ${q}$ determines which stable orientation $\vec{v_0}$ near the walls.  The orientation-dependent interaction potential near the bottom wall (\(V_b\)) has a similar form to that of \(V_t\); however, the stable degenerate configurations are shifted by an angle of \(\pi\) (i.e., the phase factor becomes \(\phi + \pi\)), and the coupling strength differs. We assume that the maximum alignment-induced angular velocity at the top wall is \(\omega_t = \omega_a\), while at the bottom wall it is \(\omega_b = \kappa \omega_a\). Here, the dimensionless factor \(\kappa\) determines the strength of the alignment interaction at the bottom wall compared to the top wall.


Among the many possible stable orientations of the self-propulsion velocity $\vec{v_0}$ near the wall, we mainly focus on three specific cases: 

(i) {\it Configuration I} considers the stable orientation where the self-propulsion velocity is perpendicular to the wall [see Fig.~\ref{F1}(a)]. Associated potential, $ V_t(\theta) = -\omega(y)\sin\theta$ and $V_b(\theta) = -\kappa V_t(\theta)$.

(ii) In {\it Configuration II}, we assume that the most stable alignments of $\vec{v_0}$ against the wall make acute angles, as depicted in Fig.~\ref{F1}(b). To be specific, here we consider the stable orientations $\vec{v_0}$ direction is tilted by an angle $\pi/4$ with respect to the normal to the wall, with the following forms of the interaction potential. For the top wall,
$V_t(\theta) =  \omega(y) \cos(4\theta) \;\; {\rm for} \;\; 0 \leq \theta \leq \pi $, otherwise  $V_t(\theta)= \omega(y)$, and for the bottom wall,
$ V_b(\theta) = \kappa\omega(y) \cos(4\theta) \;\; {\rm for} \;\; \pi \leq \theta \leq 2\pi $, otherwise $V_b(\theta)= \kappa\omega(y)$.

(iii) Finally, {\it Configuration III}, shown in Fig.~\ref{F1}(c), represents two equivalent orientations for the self-propulsion velocity near the walls, specifically at angles $\theta = 0$ and $\theta = \pi$ ($\vec{v_0}$ is parallel to the walls). We use the following form of interaction potential, \(V_t(\theta) =-\omega(y) \cos(2\theta)\) and \(V_b(\theta)= -\kappa\omega(y) \cos(2\theta)\).

For the stable configurations where the self-propulsion velocity is directed away from the walls (for example, $-\pi < \theta < 0$ for the top wall and $0 < \theta < \pi$ for the bottom wall), swimmers tend to quickly move away from the regions of alignment interactions. As a result, these alignment interactions have minimal impact on the dynamics of the active particles. Therefore, our main focus remains on the three stable configurations discussed above. Additionally, it is worth noting that when $\vec{v_0}$ makes an acute angle with respect to the walls in stable configurations, swimmers exhibit diffusion characteristics similar to those observed in {\it Configuration II}.

As the analytic solution of the Langevin equations(\ref{L1}-\ref{L3}) is a formidable task. We numerically integrated them using a standard Milstein algorithm \cite{klo} to find the particle's center of mass as a function of time.  We have taken a very small time step for numerical integration (order of $10^{-3} - 10^{-5}$) to ensure numerical stability. At the beginning,   the self-propulsion velocity orientation is taken to be uniformly distributed over the range 0 to $2\pi$.

To characterize wall particle alignment interaction-induced directional motion, we numerically estimate average drift velocity, defined as, 
\begin{eqnarray} 
	\bar{v} = \lim_{t\to\infty} \langle \left[x(t)-x(0)\right]/t \rangle.\label{drif-def1}
\end{eqnarray}
Where, $\langle \; .\; . \;. \rangle$ represents averaging over trajectories. All the particles' trajectories were allowed to evolve for long times, so that transient effects due to the initial condition completely die out. To be specific, we choose the simulation time to be the order of   $10^{3}\times 1/D_\theta$, or $10^{3}\times 1/\Omega_I$, or $10^{3}\times 1/\Omega_s$ or $10^{3}\times 1/\omega_a$, or $10^{5}$ whichever is greater. We verified that in this simulation time limit $\bar{v}$ becomes time invariant, reaching its stationary value. All the results reported in the Fig.~\ref{F2} - Fig.~\ref{F6}, obtained by averaging $10^3$ to $10^4$ trajectories in the long time limit,  where $\bar{v}$ remains constant over time.

To quantify particles' drift in the absence of external biases, we introduce the rectification power or efficiency,  
$$\eta =|\bar{v}|/v_0.  $$
It measures to what extent the self-propulsion velocity is directed toward a specific direction. Furthermore, throughout our simulations assume that \(1/\chi \gg \lambda\), which implies there will be no significant change in alignment interaction within the cutoff distance \(\lambda\). Therefore, we set \(1/\chi = 0\) to reduce the parameter space in our analysis.

Note that Eqs.~ (\ref{L1}-\ref{L3}) involve only length and time scales. In our simulations, we consider time in seconds and length in microns.  However, proper scaling of the time and length (see appendix A), reduced parameter space where independent dimensionless parameters can be chosen as,  \( Pe = {v_0}/{\sqrt{D_0 D_\theta}} \), \( \tilde{\Omega}_I = {\Omega_I}/{D_\theta} \),   \( \tilde{\Omega}_w = {\Omega_w}/{D_\theta} \),  \( \tilde{g} = {g}/{\sqrt{D_0 D_\theta}} \),   for the Couette flow described in Eq.~(\ref{Couette flow-1}) \( \tilde{u}_s = {2u_0}/{D_\theta y_L} \), and \( \tilde{\Omega}_s = {u_0}/{D_\theta y_L} \). In the figures, self-propulsion parameters and $D_0$ are  reported in terms of $Pe$, nevertheless the from the information provided in the captions, one can easily obtain exact values of all parameters.

\section{Results and discussions}\label{Results}

\begin{figure}[tp]
	\centering \includegraphics[width=8cm]{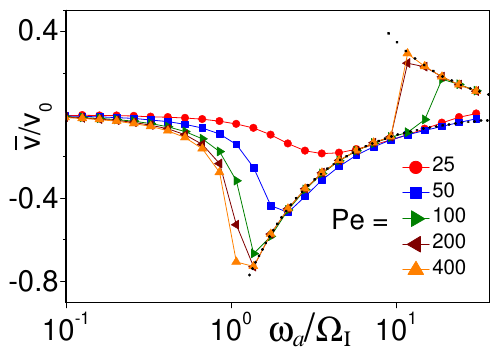}
	\caption{ (Color online)  $\overline{v}$ as a function of alignment induced coupling strength ($\omega_a$) for different  
		p\'eclet number  $Pe$ (see legends). To get different $Pe$ values, we have varied $v_0$ for $D_{\rm \theta} = 0.01 \, s^{-1}$ and $D_0 = 0.01 \, \mu m^2/s$. The dotted lines and dashed lines indicate analytical estimation based on the Eq.~(\ref{drift-approx-1a}-\ref{drift-approx-1b}) and Eq.~(\ref{drift-2nd-peak}), respectively.  Simulation parameters (unless reported otherwise in the legends):  $\Omega_I = 1 \; s^{-1}$, $\kappa = 0.125,  \; u_s = 0,\; \Omega_s = 0, \; \; g = 0,\; \lambda = 0.05\; \mu m, \; y_L = 1\; \mu m$. \label{F3}}
\end{figure}

Our numerical simulation results, presented in Fig.~\ref{F2} -- Fig.~\ref{F6}, reveal that active particles exhibit a striking directional motion, driven by their alignment interactions with wall surfaces. We systematically examine all potential stable alignments of self-propulsion velocities near the walls. Delving into each configuration, we investigate the autonomous directed motion of both chiral and achiral particles to uncover the intricate dynamics at play.

\subsection{Rectification of the motion of chiral active Janus particles for stable configuration I}

\begin{figure}[tp]
	\centering	\includegraphics[width=0.45\textwidth,height=0.14\textwidth]{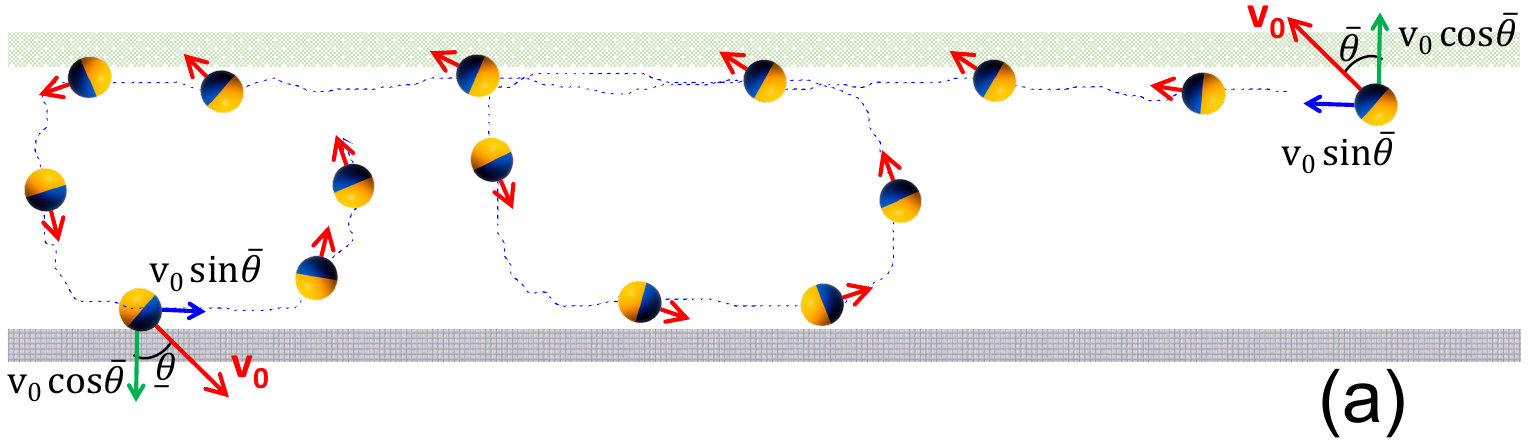}
	\centering \includegraphics[width=8cm]{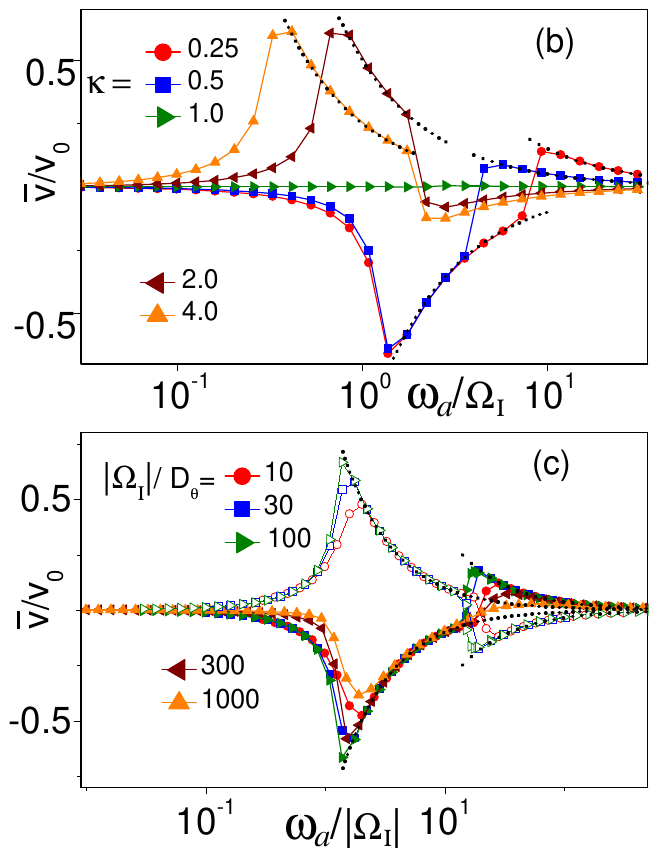}
	\caption{ (Color online) (a) The schematic depicts a typical orientation of a chiral active particle of the Janus kind, illustrating the tilting angle ($\overline{\theta}$), the direction of the self-propulsion velocity ${\vec{v_0}}$, and its normal ($v_0 \cos\overline{\theta}$) and tangential components ($v_0 \sin\overline{\theta}$). The blue dotted line represents the locus of the particle's center of mass.  (b) Average velocity $\overline{v}$ as a function of alignment-induced coupling strength ($\omega_a$) for different alignment ratios of alignment-interaction strength, $\omega_t/\omega_b = \kappa$ (see legends). (c) Similar to the panel, $\overline{v} \; vs. \; \omega_a$ for different intrinsic torques $\Omega_I$ (see legends). The plots with hollow and solid symbols, respectively, correspond to dextogyre (-ve $\Omega_I$) and levogyre (+ve $\Omega_I$) active particles. In the panels (b) and (c), the dotted lines and dashed lines indicate analytical estimation based on the Eq.~(\ref{drift-approx-1a}-\ref{drift-approx-1b}) and Eq.~(\ref{drift-2nd-peak}), respectively. Simulation parameters (unless reported otherwise in the legends): $D_{\rm \theta} = 0.01 \; s^{-1}$, $D_0 = 0.01 \; \mu m^2/s $, $v_0 = 1\; \mu m/s$, $\Omega_I = 1 \; s^{-1}$, $\kappa = 0.125,  \; u_s = 0,\; \Omega_s = 0, \; \; g = 0,\; \lambda = 0.05\; \mu m, \; y_L = 1\; \mu m$. \label{F2}}  
\end{figure}

We begin our analysis by examining the rectification of the motion of chiral active particles through  particle-wall alignments interactions when the stable self-propulsion velocity, \(\vec{v_0}\), corresponds to configuration I (see Fig.~\ref{F1}(a)). In this configuration, \({q} = 1\) and \(\phi = 0\) in Eq.~\(\ref{top_V}\). The direction-dependent alignment interaction potential reaches its minimum value when the self-propulsion velocity directs the particle toward the wall (i.e., at angles \(\theta = \pi/2\) and \(3\pi/2\) for the top and bottom walls, respectively). On the other hand, the intrinsic torque \(\Omega_I\) attempts to tilt the orientation of the self-propulsion velocity away from these potential minima. At the stable orientation of \(\vec{v_0}\), the torque resulting from alignment interactions is balanced by the intrinsic torque. For the top walls, this balance leads to the relation \(\cos\theta = -\Omega_I/\omega_a\). This chiral torque causes the self-propulsion velocity direction to tilt by an angle \(\overline{\theta}\) relative to the normal of the channel wall. The magnitude of this tilt is given by, 
\begin{eqnarray} 
	\overline{\theta} = \arcsin(\Omega_I/\omega_a) \label{tilt-angle}
\end{eqnarray}
This angle is illustrated in Fig.~\ref{F2}(a) for a levogyre active particle with respect to both the top and bottom channel walls. 
Due to this tilting of the \(\vec{v_0}\) orientation, chiral swimmers continue to slide along the channel walls, on average, with a self-propulsion velocity component of \(v_0 \cos{\overline{\theta}}\). Specifically, on the top wall, a levogyre active particle tends to slide in the negative x-direction, while on the bottom wall, it slides in the positive  direction. The particle occasionally switches between the top and bottom walls. In the strong alignment coupling limit, the average switching time can be estimated using Kramers' rate of barrier crossing~\cite{note1,note2,note3,bibitem-dsr-1,bibitem-dsr-2},
\begin{eqnarray} 
	\tau = \frac{2\pi}{\sqrt{\omega_a^2 - \Omega_I^2}} \exp{\left[ \Delta V/D_{\theta} \right]}, \label{swiching-time}
\end{eqnarray}
with the barrier height,
\begin{eqnarray} 
	\Delta V = 2\sqrt{\omega_a^2 - \Omega_I^2}+2\Omega_I \cos^{-1}\left( -\frac{\Omega_I}{\omega_a}\right)-2\pi\Omega_I. \label{swiching-barrier}
\end{eqnarray}
{\bf In Appendix B,  present details about the derivation of these equations (\ref{swiching-time}-\ref{swiching-barrier}).}  

In free space, chiral active particles follow circular paths with a radius of curvature of \(R_\Omega = v_0/\Omega_I\). However, these trajectories are significantly altered when the particles are confined within a channel narrower than \(R_\Omega\). 


\subsubsection{Autonomous directed motion induced by unequal particle-wall alignment interaction at top and bottom walls}

When the strength of alignment interactions at the top and bottom walls is equal ($\kappa = 1$), and the particle's apparent weight ($g$) is negligible compared to its self-propulsion velocity, the probabilities of the particle sliding along the top and bottom walls equilibrate. As a result, the average velocity ($\overline{v}$) of the particle becomes zero. In contrast, when $\kappa \neq 1$ or $g \neq 0$, chiral active particles exhibit spontaneous directional motion. Figure\ref{F2}-\ref{F3} illustrates $\overline{v}$ as a function of alignment interaction torques for $\kappa \neq 1$ and $g = 0$. The following key features are observed in the particle-wall alignment-induced directed autonomous motion of Janus-type active species:
\newline (i) When the alignment coupling at the top wall is significantly greater than at the bottom wall (i.e., $\kappa < 1$), in the low coupling region, a levogyre active swimmer moves spontaneously in the negative direction. However, in the strong coupling region, the direction of current reverses. Additionally, the average velocity direction changes when transitioning from $\kappa < 1$ to $\kappa > 1$.
\newline (ii) All $\overline{v}$ versus $\omega_a$ plots exhibit current reversal with both positive and negative current peaks. The peak at high $\omega_a$ is considerably weaker than the other current peak. The positions of the current peaks exhibit weak dependence on self-propulsion velocity, intrinsic torque, and rotational diffusion. In the very strong self-propulsion limit, the first peak position approaches $\omega_a \approx \Omega_I$. However, the heights of the current peaks are sensitive to these parameters.
\newline (iii) The amplitude of the autonomous drift velocity for a levogyre active particle is equal to that of a dextrogyre active particle, but in the opposite direction: 
\begin{eqnarray} 
	\overline{v}(\Omega_I) = - \overline{v}(-\Omega_I).  \label{drift-Omega}
\end{eqnarray}

{\it A. Rectification mechanism} --- To comprehend the features listed above for $\overline{v}$ as a function of $\omega_a$, we first explore the underlying mechanisms of autonomous directed motion. When the alignment coupling strength with the top and bottom boundaries of the channel differs, an active particle experiences a spatial structure with broken upside-down symmetry. As $\omega_t \neq \omega_b$, the tilting angle $\overline{\theta}$ and hence sliding time and velocity differ at the upper and lower boundaries. Thus, particles preferentially glide along a particular direction of the channel axis.   Assuming the particles spent very little time in the bulk in comparison to the channel walls, the average drift velocity of the swimmer can be expressed as, 
\begin{eqnarray} 
	\overline{v} = \frac{v_t \tau_t + v_b \tau_b}{\tau_t + \tau_b}.  \label{drift-v1}
\end{eqnarray}
Where \(\tau_t\) and \(\tau_b\) represent the sliding (or waiting) times at the top and bottom walls, respectively. We define \(\tau_t\) as the mean  time an active swimmer spends from the moment it reaches the top wall until it switches to the bottom wall. The waiting time at the bottom wall, \(\tau_b\) has a similar meaning. 
The tangential components of the self-propulsion velocities at the top and bottom walls are given by \(v_t = -v_0 \sin \overline{\theta}_t\) and \(v_b = v_0 \sin \overline{\theta}_b\). Utilizing Eq.~(\ref{tilt-angle}), one obtains, 
\begin{eqnarray} 
	\overline{\theta}_t = \arcsin(\Omega_I/\omega_a), \; \; {\rm and } \;  \overline{\theta}_b = \arcsin(\Omega_I/\kappa\omega_a)  \label{drift-v2}
\end{eqnarray} 
This relationship is  strictly restricted to the conditions \(|\Omega_I/\omega_a| \leq 1\) and \(|\Omega_I/\kappa \omega_a| \leq 1\). Thus, a simplified expression for the tangential velocity components that drags the particle to slide along the wall is,
\begin{eqnarray} 
	v_t = -v_0 \Omega_I/\omega_a, \; \;   v_b = v_0 \Omega_I/\kappa\omega_a  \label{v-components} 
\end{eqnarray} 

\begin{figure}[tp]
	\centering \includegraphics[width=8cm]{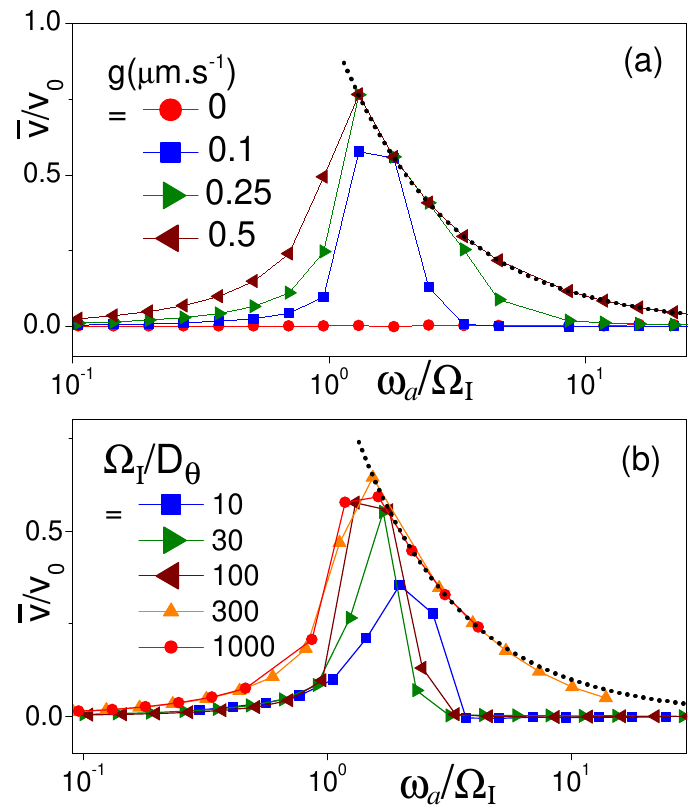}
	\centering \includegraphics[width=8cm]{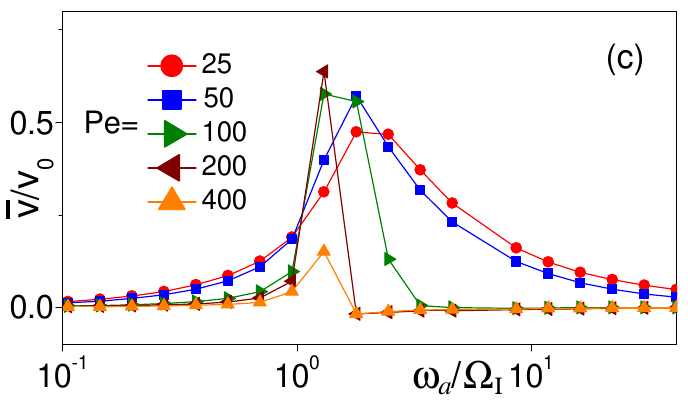}
	\caption{ (Color online) Rectification of motion of chiral active particles when upside-down symetry is broken due to particle's apparent weight. (a) $\overline{v}$ as a function of alignment induced coupling strength ($\omega_a$) for various apparent weights $g$ (see legends).  (b) $\overline{v} \; vs. \; \omega_a$ for different intrinsic torque values (\(\Omega_I\)), as shown in the legends.  (c) $\overline{v} \; vs. \; \omega_a$ for different $Pe$. To achieve different $Pe$ values, we have varied $v_0$ for $D_{\rm \theta} = 0.01 \, s^{-1}$ and $D_0 = 0.01 \, \mu m^2/s$.  In all three panels, the dotted lines represent the estimated drift velocity based on Equations (\ref{g4} - \ref{g5}). Simulation parameters (unless reported otherwise in the legends):
		$D_{\rm \theta} = 0.01 \, s^{-1}$, $D_0 = 0.01 \, \mu m^2/s$, $v_0=1\;\mu m/s$,  $\Omega_I = 1 \; s^{-1}$, $\kappa = 0,  \; u_s = 0,\; \Omega_s = 0, \; \; g = 0.1 \; \mu m/s,\; \lambda = 0.05\; \mu m, \; y_L = 1\; \mu m$
		\label{F4}}
\end{figure}

{\it B. First Current Peak in \(\overline{v}\) versus \(\omega_a\)} --- As the strength of the alignment interaction increases, active swimmers begin to experience an alignment interaction torque that attempts to keep the particle near the wall with a specific velocity \(\vec{v_0}\). This causes a levogyre active particle to slide in the negative direction on the top wall and in the reverse direction on the bottom wall. Furthermore, when \(\kappa < 1\), the swimmers spend more time near the top wall as the alignment interaction is stronger near the top wall than at the bottom wall. This results in a negative rectification velocity \(\overline{v}\). The amplitude of the current increases as the alignment interaction strengthens for $\omega_a < \Omega_I$. This is due to the fact that with increasing $\omega_a$, both the sliding velocity \(v_t\) and the time \(\tau_t\) at the top walls grow faster than the bottom wall.

The sliding speed at the top wall reaches its maximum, \(v_t = -v_0\), when \(\Omega_I = \omega_a\). At this point, the self-propulsion velocity becomes fully aligned with the channel axis. However, small directional fluctuations due to rotational diffusion may cause the particle to stray from the alignment zone. To maintain contact with the wall, the normal component of the self-propulsion velocity --- responsible for pushing the particle against the wall  --- must remain non-zero. This requires that the average orientation angle satisfies  \(\overline{\theta}_t < \pi/2\). Consequently, $\overline{v}$ cannot attend its maximum for \(\omega_a \leq \Omega_I \). 

On the other hand, for \(\kappa < 1\), if \(\omega_a\) is slightly larger than the $\Omega_I$, the torque induced by the alignment interaction at the bottom wall (\(\kappa \omega_a\)) is insufficient to cause the particle to slide along the bottom wall with significant speed and duration. Consequently, the particle's motion is rectified in the negative direction through its interaction with only the top wall. For very large \(v_0\), even when the tilting angle is slightly less than \(\pi/2\), particles remain close to the wall and slide at almost maximum speed \(v_0\). Therefore, the current peak gets very close to the condition \(\omega_a \sim \Omega_I\).  Similar reasoning for \(\kappa > 1\) suggests that a positive current peak is expected very close to the condition \(\omega_a \sim \Omega_I/\kappa\).

{\it C. Theoretical analysis of average particle current in the strong alignment interaction limit} ---  When the torque induced by particle-wall alignment (at both the upper and lower walls) is significantly greater than the intrinsic torque, the component of the self-propulsion velocity along the channel axis can be approximated by Eq.~(\ref{v-components}). Using Eq.~(\ref{v-components}) and Eq.~(\ref{drift-v1}), the average velocity for autonomous motion is given by,
\begin{eqnarray} 
	\overline{v} = \frac {v_0 \Omega_I}{\kappa\omega_a} \frac{1 - \kappa(\tau_t/\tau_b)}{1 + (\tau_t/\tau_b)}.  \label{drift-v4}
\end{eqnarray}
Further, if we assume the switching of the sliding state from the upper wall to the lower wall (and vice versa) as a barrier crossing process that follows Kramers' type rate as described in Eq.~(\ref{swiching-time}). The ratio $\tau_t/\tau_b$ can be expressed as,
\begin{eqnarray} 
	{\tau_t}/{\tau_b} = \sqrt{\frac{\kappa^2\omega_a^2-\Omega_I^2}{\omega_a^2-\Omega_I^2}} \exp\left[\frac{\Delta V_t - \Delta V_b}{D_\theta}\right]   \label{drift-v5}
\end{eqnarray} 
Here, $\Delta V_t$ and $\Delta V_b$ represent the rotational barriers set by the alignment interaction at the top and bottom wells, respectively. By utilizing Eq.~(\ref{swiching-barrier}), one can derive, 
\begin{eqnarray} 
	&& \Delta V_t - \Delta V_b = 2\left(\sqrt{\omega_a^2 - \Omega_I^2} - \sqrt{\kappa^2\omega_a^2-\Omega_I^2}\right)  \nonumber \\
	&+&2\Omega_I \left[\cos^{-1}\left( -\frac{\Omega_I}{\omega_a}\right) -\cos^{-1}\left( -\frac{\Omega_I}{\kappa\omega_a}\right)\right]. \label{swiching-barrier2}
\end{eqnarray}   
For $\omega_a \gg \{\Omega_I, \Omega_I/\kappa\}$, the ratio of average sliding time can be simplified to,   
\begin{eqnarray} 
	{\tau_t}/{\tau_b} =\kappa\exp{\left[\frac{2\omega_a}{D_\theta}(1-\kappa) \right]}  \label{drift-v5}
\end{eqnarray}
Substituting this expression into Eq.~(\ref{drift-v4}), we obtain,  
\begin{eqnarray} 
	\overline{v} = \frac {v_0 \Omega_I}{\kappa \omega_a} \frac{1 - \kappa^2 \exp{\left[\frac{2\omega_a}{D_\theta}(1-\kappa) \right]}}{1 + \kappa\exp{\left[\frac{2\omega_a}{D_\theta}(1-\kappa) \right]}}.  \label{drift-v6}
\end{eqnarray}
Although this expression is valid for the parameter regime where $\omega_a \gg \{\Omega_I, D_\theta\}$, $R_\Omega > y_L$, and $v_0$ is significantly greater than the thermal velocity ($v_{th} \sim \sqrt{D_0 \gamma_t}$), we can draw several interesting conclusions that support our numerical results.  Firstly, as expected, Eq.~(\ref{drift-v6}) confirms that $\overline{v} = 0$ when $\kappa = 1$. Secondly, when $\omega_a \gg \Omega_I$, the exponential factor $\exp\left[{2\omega_a(1 - \kappa)}/{D_\theta}\right] \gg 1$ for $\kappa < 1$, allowing us to approximate Eq.~(\ref{drift-v6}) as,  
\begin{eqnarray} 
	\overline{v} \approx - \frac {v_0 \Omega_I}{ \omega_a}. \label{drift-approx-1a}
\end{eqnarray}
Conversely, for $\kappa > 1$, the exponential factor $\exp\left[{2\omega_a(1 - \kappa)}{D_\theta}\right] \ll 1$, leading to the approximation,  
\begin{eqnarray} 
	\overline{v} \approx \frac {v_0 \Omega_I}{\kappa \omega_a}. \label{drift-approx-1b}
\end{eqnarray}
It is evident from Eq.~(\ref{drift-approx-1a}) and (\ref{drift-approx-1b}) that the direction of the average drift velocity reverses when changing from $\kappa < 1$ to $\kappa > 1$.  Furthermore, all the expressions from Eq.~(\ref{drift-v6}) to (\ref{drift-approx-1b}) indicate that $\overline{v}(\Omega_I) = -\overline{v}(-\Omega_I)$, as noted in Eq.~(\ref{drift-Omega}). For a wide range of parameter regimes, Eq. (\ref{drift-approx-1a}) and (\ref{drift-approx-1b}) align well with the simulation results presented in Fig.~\ref{F2} and Fig.~\ref{F3}.

{\it D. Current reversal and second current peak in  $\overline{v}$ versus $\omega_a$} ---
According to Eqs.~(\ref{drift-approx-1a}-\ref{drift-approx-1b}), in the large coupling limit, rectification power should monotonically decay to zero after reaching a peak around $\omega_a \sim \Omega_I$. However, simulation results reveal an unexpectedly abrupt current reversal accompanied by an additional peak in the large coupling region [see Fig.~(\ref{F2}-\ref{F3})]. This phenomenon is attributed to transient effects due to the finite simulation time. Nevertheless, this result is practically significant, as many experiments are conducted within a limited time frame, and the distance between the injecting and absorbing points is not infinitely large. Therefore, it is worthwhile to analyze transient drift in the large coupling limit.

When $\omega_a$ is appreciably greater than $\Omega_I$, the tilting angle $\overline{\theta}$ becomes very small, leading to an increase in the sliding time at both walls. As a result, particles rarely switch between the two walls. Consequently, for particle trajectories with a uniform distribution of initial positions and orientations, there is an equal probability of particles getting stuck and sliding along the top ($p_t$) and bottom ($p_b$) walls. In this parameter regime, the direction of $\vec{v_0}$ becomes almost perpendicular to the wall. Directional fluctuations due to rotational diffusion cause the particles to tend to diffuse along the wall in both backward and forward directions. For $\kappa < 1$, the tilting angle at the bottom wall is relatively larger than that at the top wall. Therefore, particles at the bottom wall exhibit more directional motion. 
As a result, the drift velocity of a levogyre active particle becomes positive, with $\overline{v} = p_t v_t + p_b v_b$. Given that $p_t = p_b = 1/2$, and applying Eq.~(\ref{v-components}), we get, 
\begin{eqnarray} 
	\overline{v} = \frac {v_0 \Omega_I}{2\kappa\omega_a} \left(1-\kappa \right) \label{drift-2nd-peak}
\end{eqnarray}
This estimate is well accord with our simulation result presented in the Fig.~(\ref{F2}-\ref{F3}) though it is limited to the large coupling limit.

{\it E. Tuning self-propulsion parameters and thermal fluctuations strength}  --- In Fig.~\ref{F3}, we explore the effects of tuning self-propulsion parameters on autonomous directional motion. The self-propulsion force ($\gamma_t v_0$) helps particles maintain contact with the walls when the tilting angle $\overline{\theta} < \pi/2$ and determines the persistence length ($l_\theta$), as well as, the radius of curvature ($R_\Omega$). As noted earlier, as long as $R_\Omega > y_L$ and $v_0 \gg v_{th}$, the $\overline{v}$ remains relatively insensitive to $v_0$. The first peak in $\overline{v}$ versus $\omega_a$ approaches $\omega_a \approx \Omega_I$ for $\kappa < 1$ and $\omega_a \approx \Omega_I/\kappa$ for $\kappa > 1$ [see Fig.~\ref{F2}(b)]. However, when $R_\Omega$ is less than the channel's cross-section, the amplitude of current is drastically suppressed as $v_0$ decreases. This is due to particles tending to rotate and form circular orbits, leading to reduced exposure to the wall where rectification occurs.

Both rotational and translational thermal diffusion disrupt particles from their stable configurations. Consequently, the amplitude of directional velocity decreases with increasing $D_0$ or $D_\theta$. Rotational fluctuations enable particles to overcome the rotational barrier set by the interplay between intrinsic torque and alignment interaction-induced torque. This effect becomes significant when the mean square deviation of orientation fluctuation about the stable configuration, $\tilde{\delta \theta} \sim \sqrt{D_\theta/\omega_a}$, becomes comparable to the angle required to escape the sliding state. Conversely, $D_0$ pushes the particles away from the wall, which diminishes the rectification effects. As a result, $\overline{v}$ begins to decrease with increasing $D_0$, particularly when $v_0$ becomes comparable to $\sqrt{\gamma_t D_0}$.

\subsubsection{Upside-down symmetry breaking by apparent weight and autonomous directed motion from particle-wall alignment interaction}

The rectification of motion for a chiral active swimmer, through alignment interactions with boundary walls, necessitates the breaking of upside-down symmetry in the channel. One option we already have discussed is symmetry breaking resulting from the unequal strength of alignment interactions at the upper and lower  channel boundaries. Another viable approach is to introduce a driving force along the transverse direction of the channel's axis. This force can naturally arise due to the gravitational drag on a particle’s apparent weight (i.e., weight minus buoyant force), denoted as \( g \). The gravitational force \( g \) pushes the particle against the bottom wall, thus breaking the upside-down symmetry in the channel. As a result, levogyre active particles spontaneously move in the positive x-direction, while dextrogyre active particles move in the negative x-direction.

Figure~\ref{F4} depicts the variation of $\overline{v}$ as a function of alignment interaction strength $\omega_a$ for different apparent weights, intrinsic torques,  self-propulsion velocity, and thermal rotational and translational noise strengths. A little apparent weight, about $10\%$ of the self-propulsion force, is sufficient to rectify the motion of a chiral active particle with more than $50\%$ efficiency. However, exploiting the ratchet effect through particle-wall alignment interaction, heavy active particles with suitably tailored self-propulsion properties can move to a particular direction along the channel axis with a velocity close to $v_0$.

Figure~\ref{F4} further reveals that for \(\Omega_I > \omega_a\), the average rectification velocity increases rapidly with rising values of \(\omega_a\). This tends to occur because the direction of \(\vec{v_0}\) begins to align, albeit at a tilting angle, against the walls as the alignment coupling \(\omega_a\) strengthens. Consequently, the probability of the particle remaining near the walls is enhanced. Additionally, particles tend to accumulate at the bottom walls due to gravitational effects. As a result, the amplitude of the positive current (for levogyre particles) gradually increases with \(\omega_a\). In contrast, once \(\Omega_I < \omega_a\), the tilting angle diminishes with an increase in \(\omega_a\), leading to a suppression of the current amplitude as \(\omega_a\) further increases, particularly beyond the threshold of \(\omega_a \sim \Omega_I\). Notably, the peak position converges to \(\omega_a \approx \Omega_I\) in the limit of high self-propulsion. 

For the parameter regime of the decaying branch of current [see Fig.~\ref{F4}(a-c)], active particles spend most of their time near the walls. Consequently, the average drift velocity, \(\overline{v}\), can be approximated as,
\begin{eqnarray}
	\overline{v} = p_t v_t + p_b v_b \label{g1}
\end{eqnarray}
Here, \(p_t\) and \(p_b\) represent the probabilities of a particle sliding near the top and bottom walls, respectively. These probabilities can be estimated using the Boltzmann distribution ($p_t/p_b=\exp{\left[-{\gamma_t g y_L}/{k_BT_{{\rm eff}}} \right]}$) along with the normalization ($p_t + p_b = 1$) condition~\cite{ginot},  
\begin{eqnarray}
	p_t = \frac{e^{-\frac{\gamma_t g y_L}{k_BT_{{\rm eff}}}}}{1+e^{-\frac{\gamma_t g y_L}{k_BT_{{\rm eff}}}}}; \;\;\;
	p_b =  \frac{1}{1+e^{-\frac{g y_L}{k_BT_{{\rm eff}}}}} \label{g2}
\end{eqnarray}
Where $T_{\text{eff}}$ is the effective temperature for a non-equilibrium system that is comprised of active particles. In the limit of fast rotational dynamics, it can be expressed as~\cite{Lowen-e2},
\begin{eqnarray}
	T_{{\rm eff}} = \frac{\gamma_t}{k_B} \left(\frac{v_0^2}{2D_\theta} +D_0 \right). \label{g3}
\end{eqnarray}
Using Eq.~(\ref{g2}) and Eq.~(\ref{v-components}) in Eq. (\ref{g1}), the average drift velocity of the active chiral particles can be written as,
\begin{eqnarray}
	\overline{v} = \frac{v_0 \Omega_I}{\omega_a} \frac{1-e^{-\frac{g y_L}{k_BT_{{\rm eff}}}}}{1+e^{-\frac{g y_L}{k_BT_{{\rm eff}}}}} \label{g4}
\end{eqnarray}
When $g y_L \gg k_B T_{\text{eff}}$, this expression can be simplified to,
\begin{eqnarray}
	\overline{v} = \frac{v_0 \Omega_I}{\omega_a} \label{g5}
\end{eqnarray}
This expression of average directed velocity, depicted in Fig.~\ref{F4}(a-c), closely corroborates our simulation results. It is important to note that the validity of Eq.~(\ref{g5}) relies on the gravitational drag (or transverse drag) being significantly large, and on the strength of thermal rotational and translational motions not being so large as to disrupt the particles' alignment near the walls.

When the self-propulsion force $\gamma_t v_0$ becomes very large, the gravitational drag resulting from the particle's apparent weight is insufficient to keep the particle near the bottom well. Consequently, the amplitude of \(\overline{v}\) becomes negligibly small [see Fig.~\ref{F4}(c)]. Conversely, the rectification effect becomes more pronounced as the radius of curvature decreases while keeping \(v_0\) fixed. This is because the particles tend to stay close to the bottom wall and occasionally exhibit orbiting motion with \(R_\Omega \gg y_L\). As a result, they do not approach the top wall, and the contribution from the first term in Eq.~(\ref{g1}) becomes almost negligible. Thus, the motion of the particles is predominantly directed by the alignment interactions at the bottom walls, enhancing the rectification effect.

\begin{figure}[tp]
	\centering \includegraphics[width=6cm]{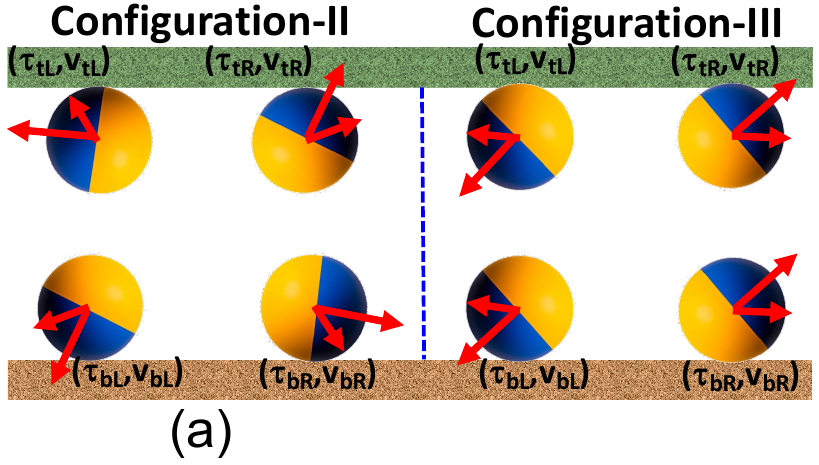}
	\centering \includegraphics[width=6cm]{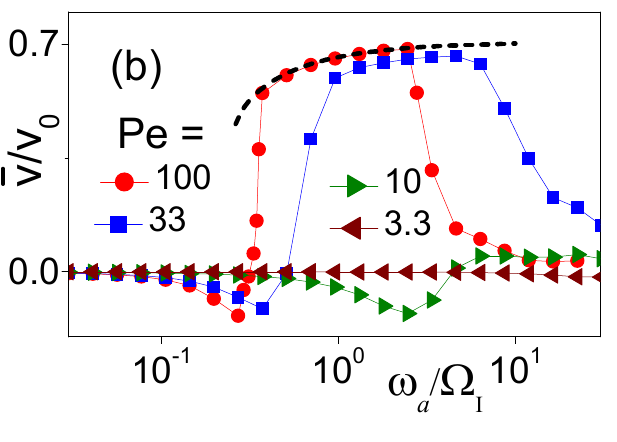}
	\centering \includegraphics[width=6cm]{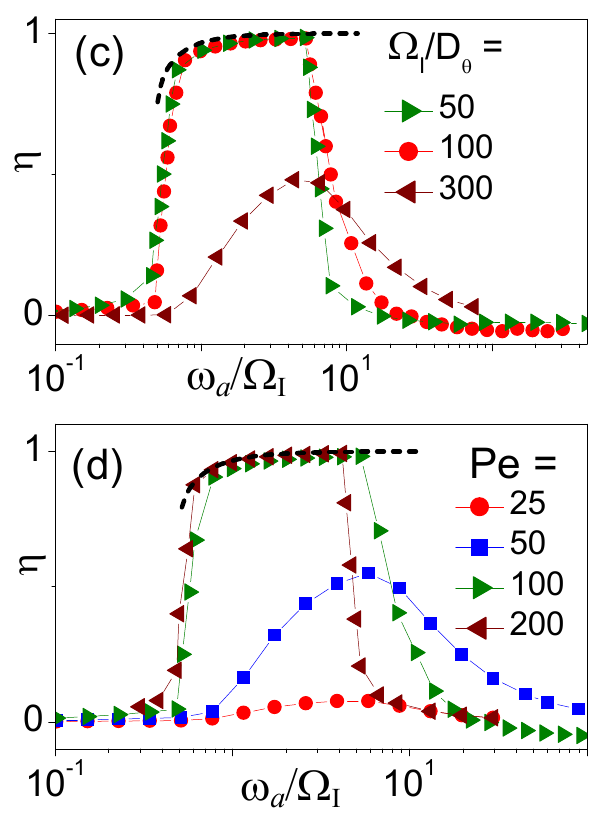}
	\caption{ (Color online) (a) Schematic diagrams illustrate the possible stable orientations of self-propulsion velocity near the walls for configurations II and III. Small arrows indicate the direction of \(\vec{v_0}\) when \(\Omega_I = 0\), while larger arrows represent the self-propulsion direction under positive chiral torque. The four states, ($\tau_{tL},v_{tL}$), ($\tau_{tR},v_{tR}$), ($\tau_{bL},v_{bL}$) and ($\tau_{bR},v_{bR}$) are characterized by their waiting times $\tau_{ij}$ ($i=t,b;\;j=L, R$) and the components of self-propulsion velocities \(v_{ij}\) along the channel axis (as explained in the text).  In panel (b), \(\overline{v}\)  is plotted against \(\omega_a/\Omega_I\) for configuration II with varying values of Pe (\(D_\theta\) values are varied for $v_0=1\; \mu m/s$ and $D_0=0.01 \;\mu m^2/s$). Dashed lines are predictions based on Eq.~(\ref{con-2_3}). Panels (c) and (d) show rectification efficiency \(\eta\) as a function of \(\omega_a/\Omega_I\) for configuration III, considering different values of  \(\Omega_I\) and \(Pe\) (refer to the legends for details). To obtain different $Pe$, here we varied $v_0$ for  $D_{\rm \theta} = 0.01 s^{-1}$, and $D_0 = 0.01 \mu m^2/s$. Dashed lines are predictions based on Eq.~(\ref{con-3_2}). Simulation parameters (unless reported otherwise in the legends):  $\Omega_I = 1.0 s^{-1}$, \; $v_0 = 1.0 \mu m/s$, \; $\kappa = 0.125, \; u_s = 0, \; \Omega_s = 0, \; \; g = 0,\; \lambda = 0.05\;\mu m, \; y_L = 1\;\mu m$. \label{F5}}
\end{figure}

\begin{figure}[tp]
	\centering \includegraphics[width=7cm]{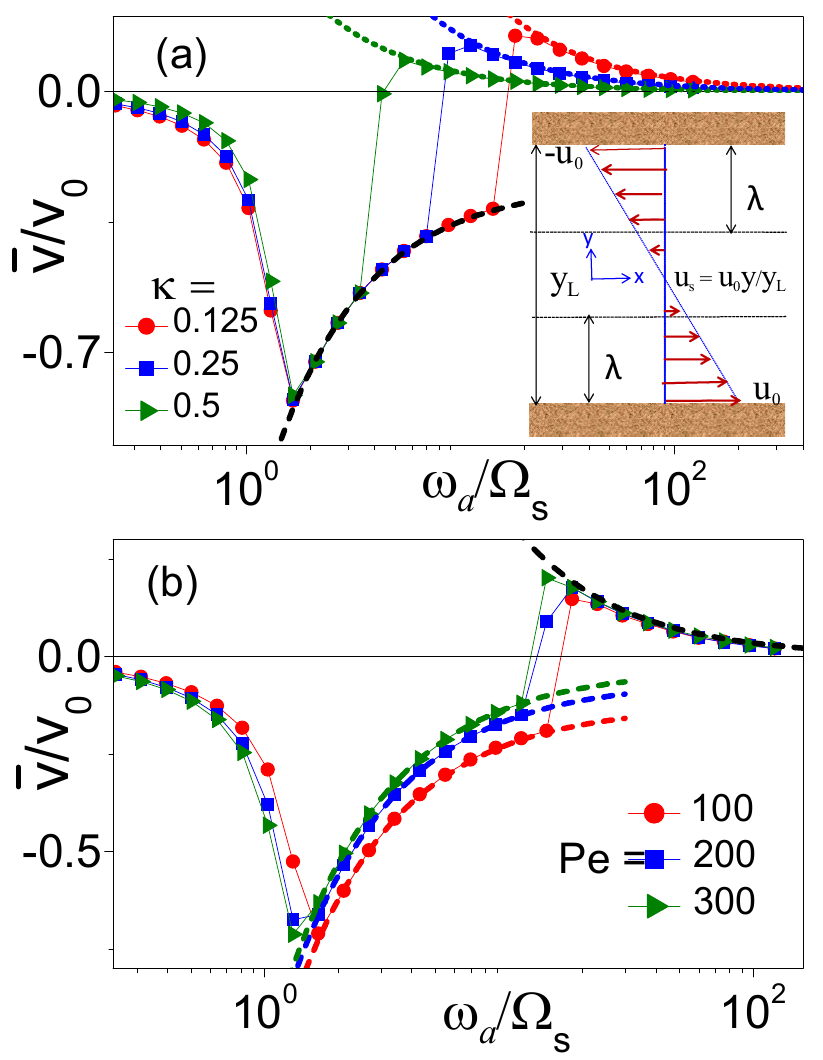}
	\caption{ (Color online) Recification of the motion of achiral active particles for configuration I. (a) The average velocity, $\overline{v}$, as a function of alignment-induced coupling strength ($\omega_a$) for different values of $\kappa$, with $u_0=0.25\;\mu m/s$ and $y_L = 1\;\mu m$. The inset illustrates a schematic of the velocity profile in a Couette flow, depicting the maximum fluid velocity ($u_0$), the channel length ($y_L$), and the cut-off distance $\lambda$ from the wall, beyond which the alignment interaction becomes zero. Panel (b) also shows $\overline{v}$ versus $\omega_a$ for various values of $Pe$ (as indicated in the legends) and with $u_0=0.125 \;\mu m/s$ and $y_L = 0.5\;\mu m$. To obtain different $Pe$, here we have varied $v_0$ for  $D_{\rm \theta} = 0.01 s^{-1}$, and $D_0 = 0.01 \mu m^2/s$.  The dashed and dotted lines represent predictions based on Eq.~(\ref{shear-v1}) and Eq.~(\ref{drift-2nd-peak-shear}), respectively. Simulation parameters (unless reported otherwise in the legends): $D_{\rm \theta} = 0.01 s^{-1}$, $D_0 = 0.01\;\mu m^2/s$, $v_0 = 1.0 \mu m/s$, $ \Omega_I = 0$, $\kappa = 0.125, g = 0, \lambda = 0.05\;\mu m,\; y_L = 1\;\mu m$. \label{F6}}
\end{figure}

\subsection{Rectification of chiral active particles for other stable particle-wall alignments configurations}
To assess the robustness of the rectification effect induced by particle-wall alignment, we investigate how variations in the alignment interaction potential impact the rectification power. Specifically, we examine two scenarios at stable states: when the self-propulsion force makes acute angles with the walls (configuration II) and when it is parallel to the walls (configuration III). Our simulation results, shown in Fig. ~\ref{F5}(b-d), demonstrate that for both configurations, the motion of chiral particles is directed, achieving rectification efficiency above 60\% for a wide range of experimentally accessible self-propulsion parameters. Further, here we focus on cases where the alignment coupling at the top wall is stronger than that at the bottom wall, which breaks the spatial inversion symmetry of the system.   

{\it Rectification in configuration II} --- Here swimmer's self-propulsion velocity ${\vec v_0}$ stabilized most  against the top (bottom) walls when $\theta = \pi/4$ and $ 3\pi/4$ ($\theta = -\pi/4$ and $ -3\pi/4$) [see Fig.~\ref{F1}(b) and Fig.~\ref{F5}(a)]. 
In this configuration,  \(\overline{v}\) as a function of coupling strength exhibits current reversals with a sharp transition occurring at high self-propulsion velocities. In the low coupling regime, a levogyre active swimmer exhibits autonomous directed motion along the negative x-direction. However, this direction reverses in strong coupling regimes. In the limit of very high coupling, the amplitude of \(\overline{v}\) approaches zero.

Figure~\ref{F5}(a) compares the stable orientation of \(\vec{v_0}\) with and without intrinsic chiral torques, represented by small and large arrows, respectively. Particles can be stabilized against the wall at four specific orientations of \(\vec{v_0}\) (indicated by large red arrows). The self-propulsion velocity directions for these states can be estimated by equating the alignment-induced torque with the intrinsic torque, i.e., \(\langle \dot{\theta} \rangle = -V'(\theta) + \Omega_I = 0\). 
This situation can be compared to the averaging out of the effects of rotational fluctuations. In the range \(0 \leq \theta \leq \pi\) (for the top wall), the equation has four roots:
$\theta = \pi/4+(1/4)\arcsin\left({\Omega_I}/{4\omega_a}\right)$, $\pi/2-(1/4)\arcsin\left({\Omega_I}/{4\omega_a}\right)$, $ 3\pi/4+(1/4)\arcsin\left({\Omega_I}/{4\omega_a}\right)$, and   $ \pi -(1/4)\arcsin\left({\Omega_I}/{4\omega_a}\right) $. 
Among these, the stable acute angles where \(V''(\theta) > 0\) are:
$\theta = \pi/4+(1/4)\arcsin\left({\Omega_I}/{4\omega_a}\right)$  and $ 3\pi/4+(1/4)\arcsin\left({\Omega_I}/{4\omega_a}\right)$. They are denoted as states (\(\tau_{tR}, v_{tR}\)) and (\(\tau_{tL}, v_{tL}\)), respectively, in Fig.~\ref{F5}(a).  Similarly, the two stable acute angles at the bottom wall are $\theta = 5\pi/4+(1/4)\arcsin\left({\Omega_I}/{4\omega_a}\right)$, and $ 7\pi/4+(1/4)\arcsin\left({\Omega_I}/{4\omega_a}\right)$. These states at the bottom wall are respectively denoted by (\(\tau_{bL}, v_{bL}\)) and (\(\tau_{bR}, v_{bR}\)) [see Fig.~\ref{F5}(a)].   

Assuming particles spend most of their time at the wall in these four orientations, the average velocity can be expressed as follows, 
\begin{eqnarray} 
	\overline{v}=\frac{\tau_{tL} v_{tL}+\tau_{tR} v_{tR}+\tau_{bL} v_{bL}+\tau_{bR} v_{bR}}{\tau_{tL}+\tau_{tR}+\tau_{bL}+\tau_{bR}}.  \label{con-2_1}
\end{eqnarray}
Here, \(\tau_{tL}\) and \(\tau_{tR}\) are the mean waiting times at stable states on the top wall, where the swimmer's self-propulsion velocity \(\vec{v_0}\) is tilted towards the left (negative direction) and the right (positive direction), respectively. The corresponding swimmer velocities at these states are denoted as \(v_{tL}\) and \(v_{tR}\)  [see Fig.~\ref{F5}(a)]. Similarly, \(\tau_{bL}\) and \(\tau_{bR}\) represent the waiting times, while \(v_{bL}\) and \(v_{bR}\) represent the velocities at the stable states on the bottom wall. When the coupling strength at the top wall is strong enough to hold the particle against it, the alignment interaction at the bottom wall is insufficient to keep the particle aligned. As a result, particles spend most of their time at the top wall. Thus, the expression in Eq.~(\ref{con-2_1}) can be approximated as,   
\begin{eqnarray} 
	\overline{v} \approx \frac{\tau_{tL} v_{tL}+\tau_{tR} v_{tR}}{\tau_{tL}+\tau_{tR}} \label{con-2_2}   
\end{eqnarray}
In the relatively large coupling region, for the stable state corresponding to the waiting time \(\tau_{tL}\), the self-propulsion is oriented at an angle $\theta = 3\pi/4+(1/4)\arcsin\left({\Omega_I}/{4\omega_a}\right)$. In contrast, for another stable  state at the top wall (with waiting time \(\tau_{tR}\)]) , the angle is  $\theta = \pi/4+(1/4)\arcsin\left({\Omega_I}/{4\omega_a}\right)$. In the first case, \(\Omega_I\) works to align the velocity vector \(\vec{v_0}\) with the channel axis, allowing the particle to diffuse easily and escape from the wall since self-propulsion holds the particle less tightly against it. While in the second case, \(\Omega_I\) directs the self-propulsion force perpendicular to the wall (see Fig.~\ref{F5}(a)), which causes the particle to be held tightly against the wall, making it difficult to escape from the state (\(\tau_{tR}\), \(v_{tR}\)). This indicates that \(\tau_{tR} \gg \tau_{tL}\). Consequently, the contribution from the state (\(\tau_{tR}\), \(v_{tR}\)) in the rectification process is significantly larger than that from the other state. Therefore, one can further approximate the average rectification velocity as, 
\begin{eqnarray} 
	\overline{v} \approx v_{tR} = v_0 \cos\left[\frac{\pi}{4}+\frac{1}{4}\arcsin\left(\frac{\Omega_I}{4\omega_a} \right)\right] \label{con-2_3}
\end{eqnarray}
This result, displayed with a dotted line in Fig.~\ref{F5}(b), is well accord our simulation data.  

{\it Rectification in configuration III} --- We now explore autonomous directed motion for a scenario where the orientation of  \(\vec{v_0}\) is stable against both walls at angles \(\theta = 0\) and \(\pi\) [as depicted in Fig.~\ref{F1}(c), also Fig.~\ref{F5}(a)]. As previously mentioned, for this stable configuration, the alignment interaction potential is given by \(V_t(\theta) =-\omega(y) \cos(2\theta)\) and \(V_b(\theta)= -\kappa\omega(y) \cos(2\theta)\). The middle panel in Fig.~\ref{F1}(c) illustrates the variation of \(V_t(\theta)\) with respect to the orientation of \(\vec{v_0}\).

In Fig.~\ref{F5}(c,d), we present the variation of the average directional velocity as a function of the alignment coupling strength for different self-propulsion parameters and chiral torques. Our simulation results indicate that the rectification power remains very close to one across a broad range of coupling strength and strong self-propulsion limits, specifically when $R_\Omega > y_L$, $v_0 \gg v_{th}$, and the strength of  $\vec{v_0}$ direction fluctuations is much smaller than the intrinsic chiral torque $\Omega_I$. Similar to configuration II, the self-propulsion velocity direction stabilizes most effectively at the four alignments [shown in Fig.~\ref{F5}(a)]. The self-propulsion velocity directions for these states can be estimated by equating the alignment-induced torque with the intrinsic torque. For those stable orientations of $\vec{v_0}$, self-propulsion drives the particles away from the wall, contributing minimally to the average velocity $\overline{v}$. Thus, we can express average velocity as,
\begin{eqnarray} 
	\overline{v} \approx \frac{\tau_{tR} v_{tR}+\tau_{bL} v_{bL}}{\tau_{tR}+\tau_{bL}} \label{con-3_1}
\end{eqnarray}
The components of self-propulsion velocity associated with stable orientations, $v_{tR} = v_0\cos\left[({1}/{2})\arcsin\left({\Omega_I}/{2\omega_a}\right)\right]$ and $v_{bL} = -v_0\cos\left[({1}/{2})\arcsin\left({\Omega_I}/{2\kappa\omega_a}\right)\right]$.
When the coupling of the particle with the top wall is significantly stronger than that with the bottom wall, i.e.,  $\tau_{tR} \gg \tau_{bL}$  , and when   $\Omega_I > 2\kappa \omega_a$, the direction of self-propulsion is primarily influenced by alignment interactions at the top wall. Under these conditions, the average directed velocity can be expressed as, 
\begin{eqnarray} 
	\overline{v} = v_0\cos\left[({1}/{2})\arcsin\left({\Omega_I}/{2\omega_a}\right)\right] \label{con-3_2}
\end{eqnarray}
This estimation, indicated by dashed lines in Fig.~\ref{F5}(c,d) is well aligned with simulation results. In the very strong alignment coupling limits, the self-propulsion velocity tends to get aligned parallel to the channel axis. As the coupling strength at the bottom wall is weaker than the top boundary, even at $\omega_a \gg \Omega_I$, on the bottom wall self-propulsion velocity direction bit more towards the wall. As a result, particles spent more time on the bottom wall ($\tau_{tR} < \tau_{bL}$) where alignment interaction directs $\vec{v_0}$ along the negative direction. This led to the current reversal.

\subsection{Rectification of motion of achiral particles}

The mechanism for the autonomous motion of active particles that we have discussed so far is limited to chiral particles. We consider cases where the upside-down symmetry of a spatial structure is broken due to unequal  strength of particle-wall alignment interactions with upper and lower boundaries, or due to gravitational drag caused by the particle's weight. A spatial structure lacking upside-down symmetry can rectify the motion of particles that exhibit orbiting behavior with a specific chirality. To achieve directed motion of an achiral active particle, the left-right spatial symmetry of the channel must be broken~\cite{our-review}. Many studies focus on this type of ratchet effect~\cite{hanggi-artf}. Another approach involves making achiral particles exhibit orbiting motion similar to chiral ones by applying an external torque. An example of this is a nonchiral active Janus particle carrying an electric charge and subjected to a magnetic field~\cite{external-torque-1}. Also,  the previous studies used shear-induced effect to sorting or population splitting of active particles~\cite{Ali-1,Ali-2}.

We investigate a mechanism for rectifying the motion of achiral particles by generating torque in their dynamics through an unbiased shear flow. Specifically, we consider an active particle that, in addition to experiencing particle-wall alignment interactions, is advected by a Couette flow --- characterized by the velocity profile depicted in the inset of Fig.~\ref{F6}(a),
\begin{eqnarray}
	u_s(y) = -\frac{2u_0 y}{y_L}  \label{Couette flow-1}
\end{eqnarray}
The flow is directed along the x-axis, with the shear gradient oriented along the y-axis. Note that the y-coordinate is delimited by $\pm y_L/2$. The swimmers experience maximum drag of \(\pm \gamma_t u_0\) at \(y = \pm y_L/2\). In addition to the dragging force, shear flow influences the dynamics of a particle, causing its self-propulsion velocity to rotate due to the local torque,   $\Omega_s = -(1/2) \vec{\nabla} \times \vec{u_s} = u_0/y_L$. As a result, in the free space, the trajectories of particles tend to bend, with the radius of curvature inversely proportional to \(\Omega_s\).

To estimate the average velocity of directed motion, we numerically solve Eq.~(\ref{L1}-\ref{L3}), taking into account shear-induced drag \(u_s\) and torque \(\Omega_s\). We consider the particle-wall alignment interaction potential corresponding to configuration I, and the upside-down symmetry of the channel is lifted by introducing different coupling strengths for the top and the bottom walls. Figure~\ref{F6} illustrates how the motion of an achiral active particle is rectified due to the interplay between shear flow and particle-wall alignment interactions.

Note that in the present set up, a particle can achieve directed motion due to both rectification of the self-propulsion motion and the obvious effect of advection. Consequently, the upper limit of the particle's average velocity for directional motion is \( v_0 + u_0 \). Furthermore, we select \( u_0 \ll v_0 \) so that the contribution of the advective effect on directional motion can be neglected. Therefore, the average velocity \( \overline{v} \) primarily results from the rectification of the self-propulsion velocity. 

Figure~\ref{F6}(a) illustrates the variation of the average directional velocity of an achiral particle as a function of alignment-induced torque for different values of $ \kappa $. All these plots display similar characteristics to those seen in chiral swimmers (refer to Fig.~\ref{F2} and \ref{F3}). In this context, the shear-induced torque, \( \Omega_s \), functions as the intrinsic torque. Equation (\ref{drift-v1}) for \( \overline{v} \) remains applicable in this case; however, the tangential components of swimmers at the top and bottom walls are given by,
\begin{eqnarray} 
	v_t = - \frac{v_0\Omega_s}{\omega_a} - u_0, \; \;   v_b =  \frac{v_0\Omega_s}{\kappa\omega_a} + u_0 \label{v-components_shear} 
\end{eqnarray}  
Following the analysis presented in Section IIIA,  Eqs.~(\ref{drift-v4}-\ref{drift-approx-1b}), we obtain a simplified expression of average velocity for directed motion when $\Omega_s/\omega_a \gg 1$, $v_0 \gg v_{th}$ and $v_0/\Omega_s > y_L$, 
\begin{eqnarray} 
	\overline{v} = - \frac{v_0 u_0}{y_L\omega_a} - u_0, \; \; {\rm for } \; \kappa<1,  \label{shear-v1}
\end{eqnarray}
and,  
\begin{eqnarray} 
	\overline{v} =  \frac{v_0 u_0}{\kappa y_L\omega_a} + u_0, \; \; {\rm for } \; \kappa>1, \label{shear-v2} 
\end{eqnarray} 
Estimates based on these equations (indicated by dashed lines) align well with the simulation results presented in Fig.~\ref{F6}. Equations (\ref{shear-v1}) and (\ref{shear-v2}), along with the simulation findings, demonstrate that the direction of the Couette flow can reverse \(\overline{v}\) direction. Additionally, the direction of rectification can also be flipped by changing \(\kappa\) from less than 1 to greater than 1.

Similar to the case of chiral particles, transient effects dominate in regions of very high coupling for a finite simulation time. This results in a reversal of the current and a current peak when the coupling strength exceeds \(\Omega_s/\kappa \). Following the same reasoning as in Eq.~(\ref{drift-2nd-peak}), we get,
\begin{eqnarray} 
	\overline{v} = \frac {v_0 u_0}{2y_L\kappa\omega_a} \left(1-\kappa \right). \label{drift-2nd-peak-shear}
\end{eqnarray}
This equation predicts that rectification power is directly proportional to the shear torque $u_0/y_L$ and inversely related to the coupling strength. Simulation results based on the Langevin equation well accord with this  prediction [see Fig.~\ref{F6}(a,b)].

\section{Conclusions}\label{Conclusions}
We present a mechanism for rectifying the motion of active particles in a narrow, structureless, straight channel, which arises from the particle-wall alignment interaction. To break the spatial symmetry of the channel and enable the underlying ratchet effect, we introduce asymmetry in the coupling strength between the two opposite boundary walls. Additionally, the spatial symmetry of the channel can be lifted due to gravitational drag for the apparent weight of the particles. 
Our study includes both chiral and achiral active particles, considering various stable alignment configurations, thus providing broad insights into controlled active matter transport.   The following is a summary of our key findings:
\newline (1) The direction and amplitude of directed autonomous motion of active particles largely depend on the orientation of self-propulsion at the stable alignment near the walls. When  self-propulsion velocity is perpendicular to the wall at the stable orientation, levogyre active particles spontaneously move in the negative direction as long as the alignment interaction with the top wall is stronger than with the bottom wall. However, in the same channel structure and with the same particle chirality, if the stable configuration has the self-propulsion velocity making an acute angle or being parallel to the channel wall, the maximum rectification power is achieved in the positive direction. Regardless of the stable velocity orientation and self-propulsion properties, the direction of the rectified motion can be reversed by making the alignment coupling with the bottom wall stronger than that with the top wall, or by inverting the chirality of the particles.  
\newline (2) The top-bottom asymmetry in the alignment coupling strength or the apparent weight due to gravitational drag can only break the upside-down spatial symmetry of the channel. This asymmetry can only rectify the motion of chiral particles that exhibit circular orbits within the channel. We show that this type of spatial structure can also be utilized to rectify the motion of achiral particles by introducing shear torque through Couette flow.   
\newline (3)   The rectification effects we demonstrated for both chiral and achiral particles are very robust in relation to particle-wall alignment interaction, self-propulsion properties, and intrinsic or shear-induced external torque. Across a wide range of parameters that are easily accessible in experiments, the rectification efficiency exceeds 60\%.
\newline (4) We observe a prolonged transient drift in the self-propulsion velocity at very high alignment coupling strengths. We illustrate that this transient behavior can be harnessed to steer the self-propulsion velocity of an active particle in the direction opposite to its long-time limit. 

In conclusion, our study presents a straightforward method for implementing the ratchet effect by exploiting the particle-wall alignment interaction to direct the motion of randomly moving active particles. Unlike previous studies, our proposed rectification setup does not necessitate any spatial periodic structures, making it simpler to implement in experiments. Furthermore, we show that the direction of rectified autonomous motion depends on the nature of the chirality. Thus, our ratcheting protocol enables the easy separation of detrogyre, levogyre, and achiral particles from their mixtures. We hope our findings will assist experimentalists in designing active Janus particles with desired transport features.        

\appendix

\section{Reduction of paramter space by rescaling time and length}\label{A}
Here, we analyse rectification of active particles' motion in reduced parameter space. To this purpose we define time and length using the units of $1/D_\theta$ and $\sqrt{D_0/D_\theta}$, respectively. Consequently,    dimension less coordinates $(\tilde{x},\tilde{y})$ and time $\tilde{t}$ are defined as,$$ \tilde{x}=\frac{x}{\sqrt{D_0/D_{\theta}}},\; \tilde{y}=\frac{y}{\sqrt{D_0/D_{\theta}}}\; {\rm and}\; \tilde{t}=\frac{t}{1/D_\theta}.  $$ The corresponding characteristic velocity is given by \(\tilde{v}_c = 1/\sqrt{D_0 D_{\theta}}\). Using these dimensionless parameters,we can express Eq.~(\ref{L1}-\ref{L3})  in the dimensionless form,
\begin{eqnarray}
	\frac{d\tilde{x}}{d\tilde{t}} &=& Pe \cos{\theta} + \tilde{u_s} \tilde{y} + \sqrt{2}\;\xi_{x}(\tilde{t})\label{Ls1} \\
	\frac{d\tilde{y}}{d\tilde{t}} &=& Pe \sin{\theta} - \tilde{g} + \sqrt{2}\;\xi_{y}(\tilde{t})\label{Ls2}\\
	\frac{d\theta}{d\tilde{t}} &=& \tilde{\Omega}_I + \tilde{\Omega}_{s} + \tilde{\Omega}_w + \sqrt{2}\;\xi_\theta(\tilde{t})\label{Ls3}
\end{eqnarray} 

In the equations presented above, the three parameters \( D_0 \), \( D_\theta \), and \( v_0 \) are consolidated into a single independent dimensionless parameter, the P\'eclet number, defined as \( Pe = \frac{v_0}{\sqrt{D_0 D_\theta}} \). Other dimensionless parameters are defined as follows [for the Couette flow described in Eq.~(\ref{Couette flow-1})]: 
\( \tilde{\Omega}_I = \frac{\Omega_I}{D_\theta} \),   \( \tilde{\Omega}_w = \frac{\Omega_w}{D_\theta} \) , \( \tilde{\Omega}_s = \frac{u_0}{D_\theta y_L} \), \( \tilde{g} = \frac{g}{\sqrt{D_0 D_\theta}} \),  \( \tilde{u}_s = \frac{2u_0}{D_\theta y_L} \).

Furthermore, the relationship \( {\tilde{\omega}_a}/{\tilde{\Omega}_I} ={\omega_a} /{\Omega_I} \) and \( {\tilde{\omega}_a}/{\tilde{\Omega}_s} ={\omega_a} /{\Omega_s} \)  holds true, and the average drift velocity in the rescaled units is given by,
\begin{eqnarray} 
	\langle \tilde{v}\rangle = \lim_{\tilde{t}\to\infty} \frac{\langle \left[\tilde{x}(\tilde{t})-\tilde{x}(0)\right]\rangle}{\tilde{t}} = \frac{\bar{v}}{\sqrt{D_0 D_\theta}}.\label{drif-def2}
\end{eqnarray}
Where $\bar{v}$ is defined by Eq.~(\ref{drif-def1}). Based on this description, all simulation results can be expressed in terms of rescaled time and length.   

\section{Derivation of  Equations (\ref{swiching-time}--\ref{swiching-barrier})}\label{B}
For configuration I, the torque experienced by the active particle due to alignment interaction and intrinsic torque is $\gamma_r(\omega_a \cos\theta + \Omega_I)$. We set the rotational frictional coefficient, $\gamma_r = 1$ which is equivalent to scaling all the parameters by $\gamma_r$. The associated interaction potential is given by, 
\begin{align}
	V(\theta)   &= -\omega_a \sin\theta - \Omega_I \theta, 
\end{align}
The maximum and minimum of the potential located at, 
\begin{align}
	\theta_{max} &= -x_c + 2\pi  \\
	\theta_{min} &= x_c
\end{align}
Where, $x_c = \cos^{-1}(-\frac{\Omega_I}{\omega_a})$. 
The associated barrier height of the potential,  
\begin{align}
	\Delta V  &=  V(\theta_{max})-V(\theta_{min}) \nonumber \\
	&= 2 (\omega_a\sin x_c+\Omega_I x_c-\pi\Omega_I)
\end{align}
Note that this expression corresponds to equation (\ref{swiching-barrier}) in Sec.~IIIA. The frequencies of the linearized potential at the bottom of the potential minima and top of the barrier are as follows:
\begin{align}
	V''(\theta_{max}) &= \omega_a \sin\theta_{max}= -\sqrt{\omega_a^2-\Omega_I^2}, \\
	V''(\theta_{min}) &= \omega_a \sin\theta_{min}=\sqrt{\omega_a^2-\Omega_I^2}.
\end{align}
Barrier crossing rate and mean exit time from the stable alignment state  can be obtained using Kramer's formula\cite{note1,note2} 
\begin{align}
	r &= \frac{\left[|V''(\theta_{max})||V''(\theta_{min})|\right]^{1/2}}{2\pi\gamma_r} \exp\left[-\frac{\{V(\theta_{max})-V(\theta_{min})\}}{k_BT/\gamma_r}\right] \\
	\tau &= \frac{1}{r} = \frac{2\pi \gamma_r}{\left[|V''(\theta_{max})||V''(\theta_{min})|\right]^{1/2}} \exp\left[\frac{\Delta V}{D_\theta}\right] 
\end{align}
Using expression of $V''(\theta_{max})$ and $V''(\theta_{min})$ we obtain Eq.~(\ref{swiching-time}) where $\gamma_r = 1$. 

\section*{Acknowledgements}
P.K.G. is supported by CSIR EMR II file no. 01/3115/23. P.B. thanks UGC,
New Delhi, India, for the award of a Senior Research Fellowship. S.N. thanks CSIR, Government of India, for the award of a Senior Research Fellowship. T.D. thanks for the award of a Humboldt Research Fellowship for Postdocs. 
\section*{Data Availability}
The data that support the findings of this study are available within the article.
\section*{Conflict of interest}
The authors have no conflicts to disclose.

\end{document}